\documentclass[
reprint,
superscriptaddress,
 amsmath,amssymb,
 aip,
 pop,
]{revtex4-1}
\usepackage{amsmath}
\usepackage[]{graphicx}

\begin{document}

\title{A comparison of weak-turbulence and PIC simulations of weak electron-beam plasma interaction}
\author{H.~Ratcliffe}\affiliation{Centre for Fusion, Space and Astrophysics, Department of Physics, University of Warwick, CV4 7AL, United Kingdom}\email{h.ratcliffe@warwick.ac.uk}
\author{ C.S. Brady} \affiliation{Centre for Fusion, Space and Astrophysics, Department of Physics, University of Warwick, CV4 7AL, United Kingdom}
\author{M. B. \surname{Che Rozenan}}\affiliation{Centre for Fusion, Space and Astrophysics, Department of Physics, University of Warwick, CV4 7AL, United Kingdom}
 \author{V.M Nakariakov}\affiliation{Centre for Fusion, Space and Astrophysics, Department of Physics, University of Warwick, CV4 7AL, United Kingdom}\affiliation{School of Space Research, Kyung Hee University, 446-701 Yongin, Gyeonggi, Korea}\affiliation{Central Astronomical Observatory of the Russian Academy of Sciences, 196140 St Petersburg, Russia }


\date{Received \today; Accepted }

\begin{abstract}
Quasilinear theory has long been used to treat the problem of a weak electron beam interacting with plasma and generating Langmuir waves. Its extension to weak-turbulence theory treats resonant interactions of these Langmuir waves with other plasma wave modes, in particular ion-sound waves. 
These are strongly damped in plasma of equal ion and electron temperatures, as sometimes seen in, for example, the solar corona and wind. Weak turbulence theory is derived in the weak damping limit, with a term describing ion-sound wave damping then added. In this paper we use the EPOCH particle-in-cell code to numerically test weak turbulence theory for a range of electron-ion temperature ratios. We find that in the cold ion limit the results agree well, but for increasing ion temperature the three-wave resonance becomes broadened in proportion to the ion-sound wave damping rate. Additionally we establish lower limits on the number of simulation particles needed to accurately reproduce the electron and wave distributions in their saturated states, and to reproduce their intermediate states and time evolution. {These results should be taken into consideration in, for example, simulations of plasma wave generation in the solar corona, of Type III solar radio bursts from the corona to the solar wind and in weak turbulence investigations of ion-acoustic lines in the ionosphere.}

\end{abstract}


\maketitle
\section{Introduction}

Fast electron beams in plasma are ubiquitous in many areas of laboratory and natural plasma physics, particularly the solar corona. In many cases, the beams are weak in comparison with the background plasma, with densities of 1 part in $10^4$ or below, in which case quasilinear theory\cite{drummond1962nucl, 1963JNuE....5..169V} can be applied to describe their evolution analytically. However this describes only the production or absorption of Langmuir waves by an electron beam. The extension of quasilinear theory to what we call weak-turbulence (WT) theory\cite{kadomtsev1965plasma,1965JAMTP...6....9L,1980MelroseBothVols,1995lnlp.book.....T} adds the effects of wave-wave interactions to the quasilinear framework in the limit that the wave modes involved are weakly damped. A damping term for the ion-sound waves is then added.


Quasilinear and weak turbulence theory have been very productively applied to simulations of weak beams {in solar and ionospheric physics. For example, the classic theory of  type III solar radio bursts\cite{1958SvA.....2..653G, 2008SoPh..253....3N,2009IAUS..257..305M}} involves the production of Langmuir waves by a streaming fast electron population, and their subsequent decay and coalescence leading to escaping radio emission\cite{2006PhRvL..96n5005L, 2014arXiv1410.2410R}. However, in the solar corona and wind where such bursts are produced, the plasma ion and electron temperatures can be comparable or even equal \cite[e.g.][]{1979JGR....84.2029G,1998JGR...103.9553N}, {and ion-sound waves will be strongly damped. Plasma wave generation may also modify streaming non-thermal electron populations in flaring coronal loops\cite{2009ApJ...707L..45H,2014PhPl...21a2903P}. In the ionosphere, plasma wave interactions may lead to naturally enhanced ion-acoustic lines seen in radar observations\cite{2010JGRA..115.2310P,2011PhPl...18e2107D}.}

Attempts to apply {kinetic} simulation methods to this problem are limited by the long timescales involved in the evolution of a weak beam (of the order $10^4$ inverse plasma frequencies for a beam of $10^{-4}$ times the density of the background), alongside the necessity to resolve the inverse plasma frequency {and the Langmuir wave dynamics. This is computationally demanding for both PIC and Vlasov methods. For PIC methods, a large number of particles-per-cell (ppc) is needed.}  For example, Dum \cite{1990JGR....95.8111D} considered relatively strong beams ($n_b/n_e\sim10^{-3}$) in a series of strict particle and fluid-particle hybrid simulations with of the order $10^5$ total particles over all species, while Kasaba et al.\cite{2001JGR...10618693K} reproduced plateau formation for a relatively strong ($2 \%$) beam with a limited number of ppc (640 over 3 species), and also briefly considered weak beams. 

Nishikawa and Cairns\cite{1991JGR....9619343N} found a decay process that did not correspond to the usual ion-sound wave scattering process, which may be due to either their dense fast beam, or the limited number of particles (36 per cell). More recently Baumg{\"a}rtel\cite{2014AnGeo..32.1025B} considered a $0.5\%$ beam with around 500 electrons per cell, and found only limited evidence for weakly non-linear wave-wave coupling. Ganse et al.\cite{2012SoPh..280..551G} considered similar beam densities without ion dynamics, and therefore could not reproduce the strong Langmuir wave scattering predicted by weak turbulence. {Vlasov simulations of weak beam-plasma interactions are less common. For example, Daldorff et al.\cite{2011PhPl...18e2107D} considered ionospheric electron beams modifying ion-acoustic lines,  but were forced to use unphysical parameters such as a 1\% beam density for computational reasons, altering the processes of Langmuir wave generation and scattering. Henri et al.\cite{2010JGRA..115.6106H} considered finite Langmuir wave-packets and concluded that for sufficiently fast wave growth ion-sound wave damping could be overcome even in plasma of equal ion and electron temperature, and resonant wave-wave interactions would proceed as usual.}

In this paper we consider the beam-plasma interaction for a weak ($n_b/n_e=10^{-4}$) beam {using a PIC code}, and derive for the first time, the Langmuir wave spectral energy densities produced. We compare these directly with the predictions of weak turbulence using a previously developed code\cite{2002PhRvE..65f6408K}. We investigate the beam-plasma interaction for a range of ion temperatures, both in the cold ion regime where weak turbulence theory applies, and the hot ion case where it may be expected to breakdown. 

The paper is organized as follows. 
In Section \ref{sec:QL_WT} we introduce quasilinear and weak turbulence theory and their governing equations. In Section \ref{sec:PIC_decrip} we describe the PIC code setup and establish convergence of the results. In Section \ref{sec:compare} we compare the PIC code and WT results for cold ions and show good agreement, then treat the problem for ion-electron temperature ratios from 0.01 and 2, and derive a simple model of resonance broadening which can explain the observed discrepancies in the hotter cases. Section \ref{sec:concl} summarizes our findings.

\section{Quasilinear and weak turbulence theory}\label{sec:QL_WT}

The problem we consider is that of an approximately Maxwellian beam of fast electrons travelling in a collisionless plasma with a weak background magnetic field. {In most cases mentioned in the introduction, a 1-D model is used as} the electrons are assumed to be effectively constrained to the field lines\cite{1990SoPh..130..201M,1994PhPl....1.1821D}. The behaviour of Langmuir waves is key in determining the level and spectrum of this radio emission. In particular, initial generation of a beam parallel Langmuir wave population is expected to produce a secondary daughter wave population approximately anti-parallel, and the position and width of this is key for generation of emission at twice the plasma frequency by wave coalescence.
We describe the electrons using their distribution function $f_{\mathrm{v}}(t)$ [cm$^{-4}$ s],
where \begin{equation}\int_{-\infty}^{\infty} f_{\mathrm{v}}(t)\, {\mathrm d}{\mathrm v}=n_e +n_b,\end{equation} with $n_e$ the density of the background plasma, and $n_b$ that of non-thermal electrons,
and the Langmuir waves by their spectral energy density as a function of wavenumber $k$, $W(k,t)$ [erg cm$^{-2}$],
where
\begin{equation}\int_{-k_{De}}^{k_{De}} W_k(t)\, {\mathrm d} k=E_L\end{equation}
is the total energy density of the waves in erg cm$^{-3}$ and $k_{De}=2\pi/\lambda_{De}= 2\pi\omega_{pe}/{\mathrm v}_{Te}$ with
${\mathrm v}_{Te}=\sqrt{k_B T_e/m_e}$ the electron thermal velocity and $\omega_{pe}=\sqrt{4\pi n_e e^2/m_e}$
the plasma frequency, and $m_e, e$ the mass and charge of an electron respectively. CGS units are used to agree with e.g. Tsytovich\cite{1995lnlp.book.....T}.

The equation for the evolution of the electron distribution\cite{drummond1962nucl, 1963JNuE....5..169V} is (with explicit time dependence left out for clarity)
 \begin{equation}\label{eqn:4ql1}
\frac{\partial f_{\mathrm{v}}}{\partial t}= \frac{4\pi^2
e^2}{m_e^2}\frac{\partial}{\partial {\mathrm v}}\left( \frac{W_k}{{\mathrm
v}}\frac{\partial f_{\mathrm{v}}}{\partial {\mathrm v}}\right) \end{equation}
while the Langmuir wave evolution\cite{drummond1962nucl, 1963JNuE....5..169V,1970SvA....14...47Z} is given by

\begin{equation}\label{eqn:4ql2}
\frac{\partial W_k}{\partial t} =\frac{\omega_{pe}^3 m_e}{4\pi n_e}{\mathrm v}\ln\left(\frac{\mathrm v}{\mathrm v_{Te}}\right)f_{\mathrm{v}}
+\frac{\pi\omega_{pe}^3}{n_ek^2}W_k\frac{\partial f_{\mathrm{v}}}{\partial {\mathrm
v}}. \end{equation} 
The first RHS term here is spontaneous Langmuir wave generation, the equivalent term for electrons being negligible and thus omitted. The second RHS term is known as non-linear Landau damping. The interaction described is resonant, requiring the Langmuir wave phase velocity to match the electron velocity, or $\omega_{pe}=k \mathrm{v}$ using the common approximation that the Langmuir wave frequency is approximately the plasma frequency\cite{1963JNuE....5..169V}. 

The QL equations describe the interaction of beam of electrons with plasma and the subsequent generation of Langmuir waves. However, it is known that these Langmuir waves are subject to further instabilities, most importantly the Langmuir decay instability (LDI), whereby a Langmuir wave (L) decays to a second Langmuir wave plus an ion-sound wave (s), $L\rightleftarrows L' + s$.  Additionally, Langmuir waves may be scattered by plasma ions, but for the parameters used here this occurs far slower than the LDI\cite{2000PhPl....7.4901C} and is therefore omitted. {The following equations were originally derived in the limit of weak damping of all wave modes involved, therefore in particular in the limit $T_i \ll T_e$.} The strong ion-sound wave damping present in plasma of {similar or} equal ion and electron temperature is described by adding a damping term to the equation describing the ion-sound wave evolution, and is assumed not to affect the three-wave resonance directly.\cite{1982PhFl...25..392H}
 
 The general expression describing the LDI process is standard \cite[e.g.][]{1980MelroseBothVols, 1995lnlp.book.....T}, and is, assuming 1-D wave distributions
\begin{align}\label{eqn:4ql_sSrc}
\frac{\partial{W_k}}{{\partial t}}=&\alpha_s\omega_{k} \int \mathrm{d}k'   \omega_{k'}^s 
\notag\\&\Biggl[ \left(
\frac{W_{k-k'}}{\omega_{k-k'}}\frac{W^s_{k'}}{\omega^s_{k'}}-
\frac{W_k}{\omega_k}\left(\frac{W_{k-k'}}{\omega_{k-k'}}+
\frac{W^s_{k'}}{\omega^s_{k'}}\right)\right)\times \notag\\&\delta (\omega_{k}-\omega_{k-k'}-\omega^s_{k'})\Biggr.
\notag \\&
-\left.
\left(
\frac{W_{k-k'}}{\omega_{k-k'}}\frac{W^s_{k'}}{\omega^s_{k'}}-
\frac{W_k}{\omega_k}\left(\frac{W_{k-k'}}{\omega_{k-k'}}-
\frac{W^s_{k'}}{\omega^s_{k'}}\right)\right)\right. \Biggl.\times \notag\\&\delta(\omega_{k}-\omega_{k-k'}+\omega^s_{k'}) \Biggr] ,
\end{align}
where $W^s_{k'}, \omega_{k'}^s$ are the spectral energy density and frequency of ion-sound waves, given by 
\begin{equation}\label{eq:is_disp}\omega_{k'}^s= \frac{k' {\mathrm v}_s}{\sqrt{1+ k'\lambda_{De}}} \simeq k' {\mathrm v}_s \end{equation}
 with ${\mathrm v}_s=\sqrt{k_BT_e(1+3T_i/T_e)/M_i}$ the sound speed, and the constant is
\begin{equation}
\alpha_s=\frac{\pi \omega^2_{pe}(1+3T_i/T_e)}{4n_ek_B T_e}.
\end{equation}

Similarly the sound waves obey \begin{align}\label{eqn:4ql_s}
\frac{\partial W^s_k}{\partial t}=&-\gamma^s_k
W^s_k-\alpha_s ({\omega^s_k})^2\int \mathrm{d}k'  \notag\\&
\Biggl[
\frac{W_{k-k'}}{\omega_{k-k'}}\frac{W^s_k}{\omega^s_{k}}-
\frac{W_{k'}}{\omega_{k'}}\left(\frac{W_{k-k'}}{\omega_{k-k'}}+
\frac{W^s_k}{\omega^s_{k}}\right)\Biggr]\times\notag\\&
\delta(\omega_{k'}-\omega_{k-k'}-\omega^s_k).
\end{align}
The first term here is Landau damping of the waves,
with coefficient
\begin{align}\label{eq:isdamp}
&\gamma_S(k)=\sqrt{\frac{\pi}{8}}\omega^s_k\times\notag\\&\left[\frac{{\mathrm v}_s}{{\mathrm v}_{Te}}+\left(\frac{{\mathrm v}_s}{{\mathrm v}_{Ti}}\right)^3\exp\left[- \left(\frac{{\mathrm v}_s^2}{2{\mathrm v}_{Ti}^2 }\right) \right]\right].
\end{align}

Equations \ref{eqn:4ql1} to \ref{eq:isdamp} constitute what we call the weak-turbulence (WT) theory. Simultaneous solution of energy (frequency) and momentum (wavenumber) conservation wight he appropriate dispersion relations shows that for the process $L \rightarrow L' +s$, $k-k' \simeq - k$, and $k' \simeq 2 k$, i.e the initial Langmuir wave is approximately backscattered. More precisely, we have $k-k' = - k + \Delta k$ with the small decrement \begin{equation}\label{eq:decd}\Delta k= 2 \sqrt{m_e/M_i} \sqrt{(1+3T_i/T_e)}/(3\lambda_{De}).\end{equation}

\section{PIC code setup and convergence tests}\label{sec:PIC_decrip}

We use EPOCH1D\cite{BradyandArber2011},
a Birdsall and Langdon type PIC code\cite{birdsall2004plasma} using Villasenor and Buneman current deposition \cite{Villasenor1992306}. This solves Maxwell's equations combined with the equations of motion for charged particles in an EM field to provide a direct simulation of collisionless plasmas.

We assume hydrogen plasma with a plasma frequency of 200~MHz, and set the electron ion mass ratio as 1:1836.2. The {background plasma is Maxwellian with an} electron temperature of 2~MK, and an ion temperature varying from 0.01 times to double this. {We note that non-Maxwellian backgrounds may occur in some situations of interest, such as the kappa distributions seen in some regions of the solar wind\cite{1997GeoRL..24.1151M}, and may affect the plasma dispersion, but these are not considered here.} The simulation box length is 1024 $\lambda_{De}$ and we use periodic boundary conditions. In order to resolve the Langmuir wave dynamics, we output the electric field information every $0.3\omega_{pe}^{-1}$. The number of simulation particles per cell (ppc) varies from 400 to 24000 for each species, for 3 species (background electrons, beam electrons and ions), with the majority of simulations using 12000 each. We inject a Maxwellian beam with drift velocity $10 \mathrm{v}_{Te}$ and temperature 2~MK at time $t=0$ uniformly along the box.  

The simulations have total run time of between one and three times $ \tau_{ql}$, where  \begin{equation}\label{eq:t/l} \tau_{ql}= \frac{n_e}{n_b\omega_{pe}}\end{equation} is the ``quasilinear time'' dictating the timescale for beam-plasma interaction. The beam density is $n_b/n_e = 10^{-4}$. The quasilinear theory described above applies only in the weak beam limit, where the beam electrons do not modify the Langmuir wave dispersion. An approximate analytical limit for $n_b$ was found by Muschietti and Dum\cite{1991PhFlB...3.1968M}, which for the beam temperature used here is $n_b/n_e  = 4\times 10^{-5}$. Thus we first confirm that the beam electrons are not affecting the plasma wave dispersion.

\begin{figure}[]
\centering 
\includegraphics[scale=0.4]{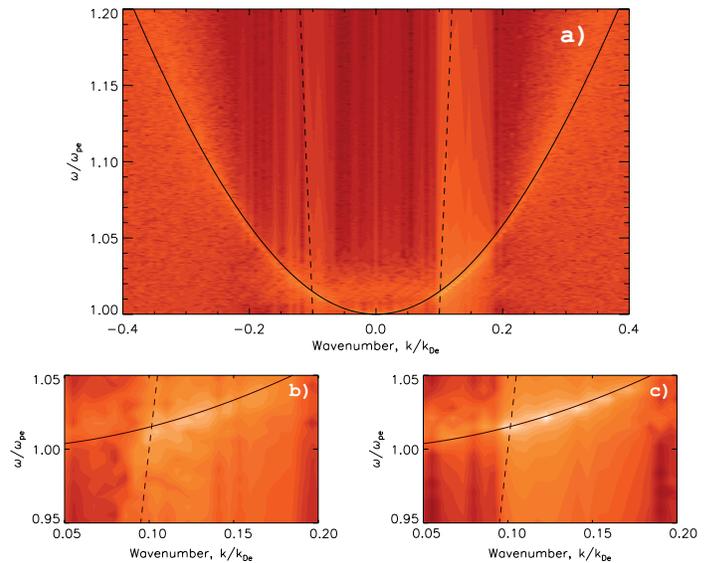}

\caption{An example Fourier transform of the electric field for a PIC code run with 12,000ppc. Panel a) shows the entire curve for a beam density of $n_b/n_e=10^{-4}$. Beam parallel waves appear at positive wavenumber, antiparallel at negative. The smaller panels show the detail of the beam parallel waves for b) $n_b/n_e=10^{-3}$ and c) $10^{-4}$. The solid black curve shows the Bohm-Gross dispersion relation for Langmuir waves (Equation \ref{eq:bg}), the dashed that for the ``beam-mode'' $ \omega = k \mathrm{v}_b $. {The color represents power on a log scale from dark to light}.\label{fig:beam_mode}}
\end{figure}

In Figure \ref{fig:beam_mode} we show an example of the Langmuir wave dispersion curve obtained from our simulations, overlaid with the Bohm-Gross relation
\begin{equation}\label{eq:bg} \omega = (\omega_{pe}^2 + 3 \mathrm{v}_{Te}^2 k^2)^{1/2}   \end{equation}
and that for the ``beam mode'' $ \omega = k  \mathrm{v}_b $ found by Willes and Cairns\cite{2000PhPl....7.3167W} for strong beams. A break in the dispersion relation is visible where these intersect in panel b) which shows a beam 10 times denser than our main simulations, with significant power visible along the dashed curve rather than the solid one below $k/k_{De} =0.1$. This arises due to the influence of beam electrons on the plasma dispersion. For the beam density used in the main simulations shown in panels a) and c) no prominent beam mode is visible {and we conclude that the Langmuir wave dispersion is well described by the Bohm-Gross relation.}

\begin{figure}
\centering 
\includegraphics[scale=0.5]{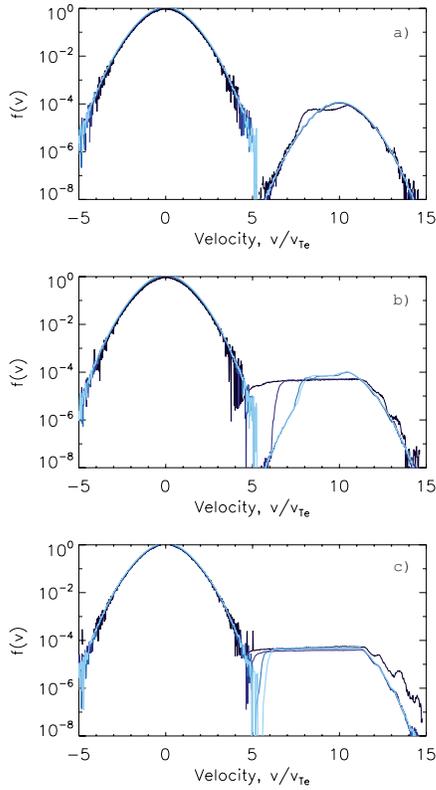}

\caption{Electron distribution functions $f(v)$, normalized to a peak value of 1, for $n_p$ = 400, 4000, 12000, 24000 from dark to light blue respectively, at times of a) $0.125 \tau_{ql}$, b) $0.5 \tau_{ql}$ and c) $ \tau_{ql}$.\label{fig:convergenceEl}}
\end{figure}

\begin{figure}
\centering 
\includegraphics[scale=0.5]{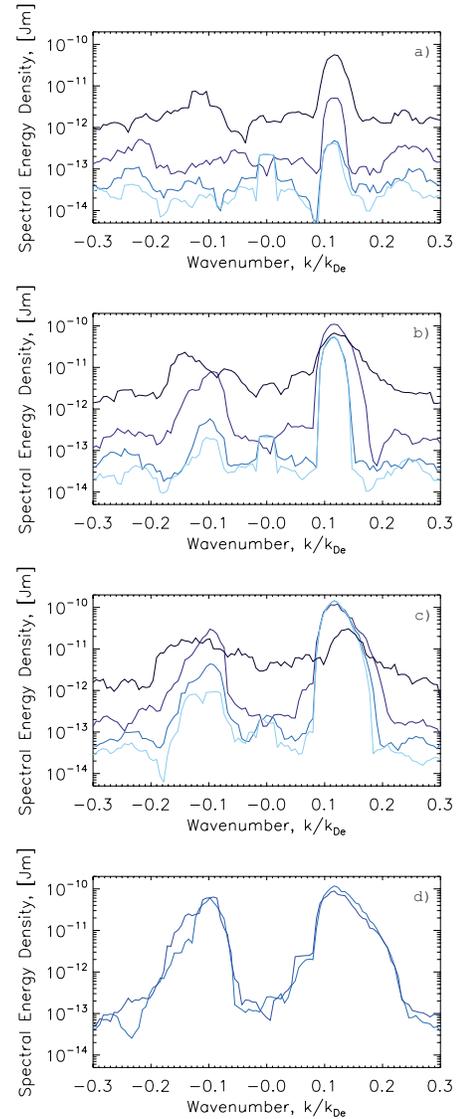}
\caption{Derived spectral energy densities of Langmuir waves against wavenumber for cold plasma ions ($T_i=0.1 T_e$), for $n_p$ = 400, 4000, 12000, 24000 from dark to light blue respectively, at times of  a) $0.125 \tau_{ql}$, b) $0.5 \tau_{ql}$, c) $ \tau_{ql}$ and d) $2 \tau_{ql}$ (6 and 12000 ppc).
Curves are smoothed over 0.025 $k/k_{De}$. \label{fig:convergence}}
\end{figure}

From  the electric field we calculate a Langmuir wave spectral energy density as follows. We take a series of 2-D windowed Fourier transforms centred on the desired time with fixed characteristic width of $0.1\tau_{ql}$. Each resulting Fourier transform is mod-squared to give energy density, and then integrated over a $\pm 2\%$ frequency band around the Langmuir wave dispersion curve, producing a spectral energy density as function of wave-number $k$. With appropriate constants restored, these are directly comparable to the wave spectral energy densities in Equation \ref{eqn:4ql2}. For the remainder of this paper we present all quantities in SI units. 

Next, we establish the convergence of the PIC code results by a series of test runs with increasing number of particles per cell per species, $n_p$. Figure \ref{fig:convergenceEl} shows the electron distribution functions for $n_p$ from 400 to 24000 ppc. In all cases the expected quasilinear plateau formation is seen, with timescale given roughly by the quasilinear time (Equation \ref{eq:t/l}).

The effects of the decreasing noise levels are immediately obvious: for small particle number the plateau formation proceeds significantly faster. However, the final results are quite similar in all but the smallest particle number case. 
The small high energy tail in the distribution function produced in the low ppc result and not in the higher number simulations should be noted, as is appears to be erroneous. In some cases, prominent tails are predicted by the WT theory\cite{2009AIPC.1188...59Y}, and this result implies high particle numbers are needed to properly reproduce this in PIC simulations.  The stepped structure in this tail arises from scattering of Langmuir waves to smaller wavenumber, as described by Dum\cite{1990JGR....95.8111D}, and is stronger in the low particle number case due to higher noise levels.

Figure \ref{fig:convergence} shows the derived spectral energy density of Langmuir waves for a simulation with cold ions ($T_i = 0.1 T_e$) and $n_p$ between 400 and 24000 ppc.  
As expected, the observed noise levels are proportional to $\sqrt{n_p} $. The increased noise associated with smaller numbers of particles causes earlier growth of  both the primary (positive wavenumber) and backscattered (negative wavenumber) Langmuir waves, but the final level of primary waves is unaffected in all but the lowest ppc case. The longer time data for 6 and 12000 ppc shows that at saturation the secondary peak height and width are also very similar. 

This suggests that around 4000 ppc is adequate to reproduce the long time dynamics of waves, but at least 12000 ppc is required to see the time evolution. For particle numbers below about 1000ppc the Langmuir wave backscatter is strongly affected in rate and spectrum.  However, 400ppc appears sufficient to reproduce general features of quasilinear plateau formation and primary Langmuir wave generation.

\begin{figure}[]
\centering 
\includegraphics[scale=0.4]{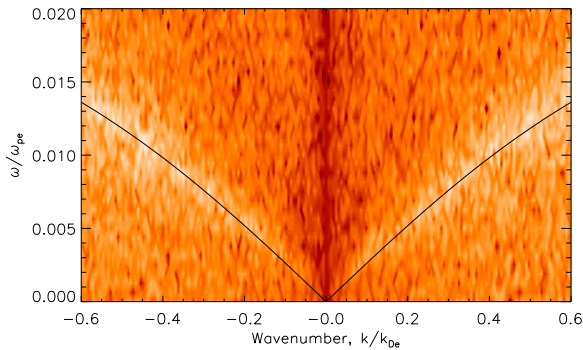}

\caption{FT of the electric field for cold ions with 12000 ppc, showing a clearly visible ion-sound mode. The solid line is the predicted dispersion relation given by Equation \ref{eq:is_disp}. {The colors show power on a log scale from dark to light.}\label{fig:ion_mode}}
\end{figure}

Finally, Figure \ref{fig:ion_mode} shows the  Fourier transform of the electric field at low frequencies, overlaid with the ion-sound wave dispersion relation from Equation \ref{eq:is_disp}. The mode is clearly visible, with the expected dispersion. Presence of this mode confirms that Langmuir wave backscattering due to wave-wave processes can occur and be resolved in these simulations.  {However their enhancement at wavenumbers around $0.2$ to $0.3 k_{De}$ due to wave-wave scattering is not distinguishable here due to noise.}

\section{Comparison between PIC and quasilinear simulations}\label{sec:compare}

\subsection{General behaviour}

\begin{figure*}
\centering 
\includegraphics[scale=0.5]{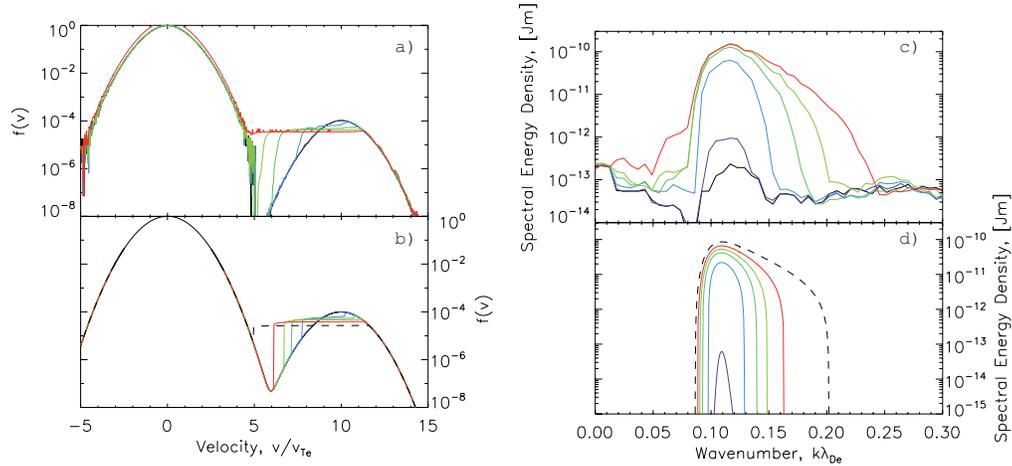}

\caption{ Left column: a) PIC code and b) WT code electron distribution functions, normalised to a peak value of 1. Right column: c) PIC and d) WT Langmuir wave spectral energy densities for positive wavenumbers. The colored lines (black to red) show 6 equally spaced times from 0.05 to 1.2 $\tau_{ql}$ respectively. The dashed line in the QL plots shows the result at $t= 3 \tau_{ql}$. The ion temperature is $0.1 T_e$. \label{fig:electronDist}}
\end{figure*}

Figure \ref{fig:electronDist} shows the electron distributions and Langmuir wave spectral energy densities in the cold ion case from the high-particle-number PIC and the WT simulations. The classic quasilinear relaxation and plateau formation is seen, and is similar in both cases. The PIC results for the electron distribution are seen to pre-empt the WT results, with the plateau widening much faster. However the final states are very similar. The Langmuir wave results also show this, producing slightly faster growth and a wider peak at all times, although the peak height and shape is very similar, and the location is identical. The dashed curves in panels b) and d) show the WT results at a later time when the plateau formation is close to saturation, which are close to the PIC result at $\tau_{ql}$ given by the red lines. This faster evolution is inferred to be produced by the PIC code noise level, which is many orders of magnitude above the thermal level of Langmuir waves. The waves therefore grow from a higher base, reaching a given level sooner, but saturate at the same level dictated by the beam density.

\subsection{Results for varying ion temperature}

\begin{table}
\begin{tabular}{ l ccc}
\hline
Run &$T_i/T_e$ & $\Delta k/k_{De}$ & $\gamma_s/\omega_s$\\
\hline
Very Cold &$0.01$&$0.024 $&$0.029 $\\
Equal &$1$&$0.046 $&$0.24 $\\
Hot &$2$&$0.061 $&$ 0.32$\\
\hline
\end{tabular}
\caption{The Langmuir wavenumber decrement (Equation \ref{eq:decd}), and the {predicted} ion-sound wave damping rate in inverse wave periods (Equation \ref{eq:isdamp}) for three electron-ion temperature ratios.\label{tab:params}}
\end{table}

Next, we compare the results for fixed electron temperature and three ion-electron temperature ratios: colder ions where WT theory works well, as shown in Figure \ref{fig:electronDist}, equal electron and ion temperatures, and a case with the ion temperature double the electron temperature. The former was performed using 10,000 ppc to allow longer runtime, the latter two with 12,000ppc. 

The former two had box length 1024 $\lambda_{De}$ as above, giving $k$-resolution of $0.006 k_{De}$. The primary peak occurs at around 0.11 $k_{De}$, and Table \ref{tab:params} shows the expected decrement in Langmuir wavenumber (Equation \ref{eq:decd}). For the hot ion case, significant wave generation is expected at wavenumbers around $0.05 k_{De}$ so to ensure these can be well resolved, a 2048 $\lambda_{De}$ box length was used in this case. Table \ref{tab:params} also shows the ion-sound wave damping rate {predicted by} Equation \ref{eq:isdamp} in inverse wave periods for the three cases. This is seen to be weak in the very cold case, but approaches critical damping for the hotter ions. 

\begin{figure}[]
\centering
\includegraphics[scale=0.5]{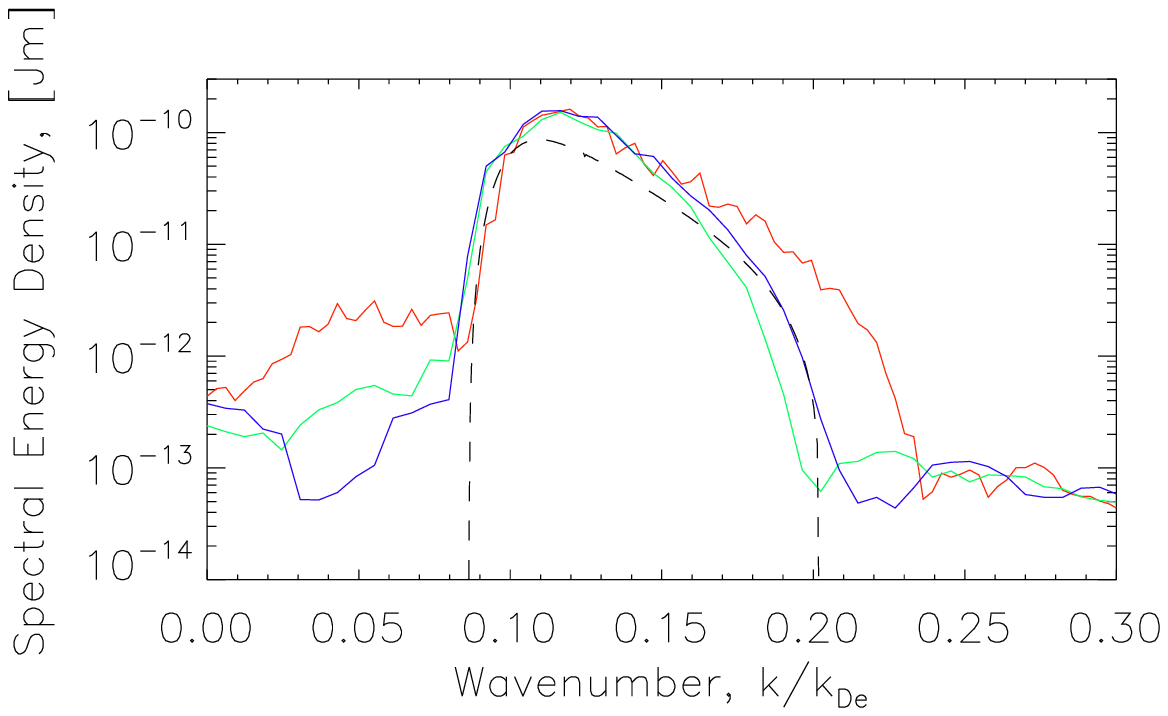}

\caption{Spectral energy densities of beam parallel primary Langmuir waves from PIC code for cold ions ($T_i=0.01 T_e$, blue), equal ($T_i=T_e$, green) and hot ions ($T_i=2T_e$, red), and the WT result (dashed line) which is independent of ion temperature. The PIC results are shown for $\tau_{ql}$, while the WT result is at $3 \tau_{ql}$.\label{fig:FwEnDens}}
\end{figure}

Figure \ref{fig:FwEnDens} shows the primary Langmuir wave peak for the three cases, together with the WT result. As established in the previous section, the PIC results pre-empt the WT code and so we plot the latter at a later time. The peak is seen to be very similar for the three cases. Small differences are visible, such as the slightly increased peak height relative to the quasilinear results, and the increased width in the hot ion case. Due to the stochastic nature of PIC simulations, it is not possible to definitively state whether these are due to simulation noise, uncertainties in the distribution reconstruction, or actual differences in the underlying physics. 

\begin{figure}[]
\centering
\includegraphics[scale=0.5]{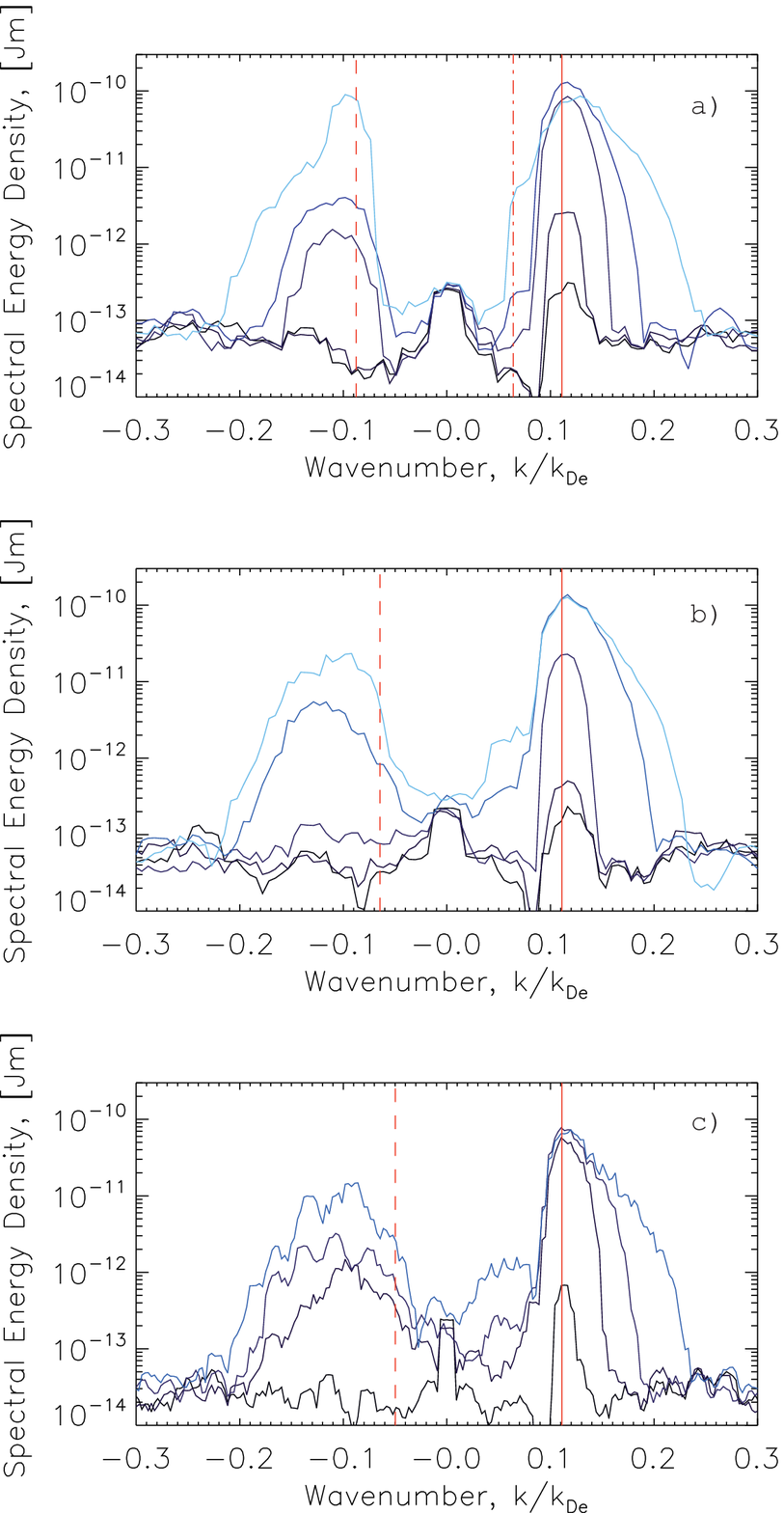}

\caption{Spectral energy densities of Langmuir waves from PIC simulations for $T_i/T_e$ of a) 0.01, b) 1 and c) 2. Blue lines from dark to light respectively show times of 0, 0.25, 0.5, 1 and 1.5 $\tau_{ql}$ (first 4 only for the hot ions case). The red vertical lines show the predicted peak positions from WT theory: solid for the initial peak, dashed for the first backscatter and dot-dashed for the second backscatter (cold ions only). \label{fig:AllEnDens}}
\end{figure}

Figure \ref{fig:AllEnDens} shows the evolution of the Langmuir waves over time from the PIC code in the three cases. For the cold ion case a secondary peak is strongly visible at the later times and has a maximum where predicted by the WT theory, that is approximately equal and opposite the forwards peak, with small decrement $\Delta k$ given by Equation \ref{eq:decd} and in Table \ref{tab:params}. The peak width also matches the WT prediction. A second scattered peak is also visible at small positive wavenumbers.

\begin{figure}[]
\centering
\includegraphics[scale=0.5]{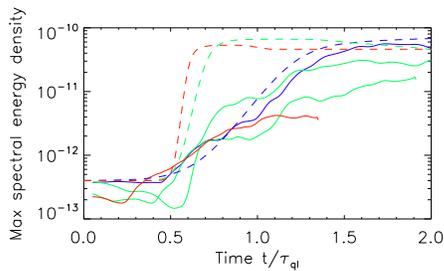}

\caption{The maximum of the secondary (negative wavenumber) peak spectral energy density as function of time. Solid lines are the PIC results, dashed lines the weak-turbulence code. The colors show ion temperature: $T_i=2 T_e$ (red), $T_i= T_e$ (green) and $T_i=0.01 T_e$ (blue). For equal temperatures we show 2 independent runs of the PIC code with different seeds for the random number generator, and hence different realisations of the initial conditions. The weak-turbulence results are shifted in time such that they exceed the PIC noise level at 0.5 $\tau_{ql}$, and have added noise baseline of $4\times10^{-13}$.  \label{fig:GrowthRates}}
\end{figure}

Figure \ref{fig:GrowthRates} shows the time evolution of the daughter peak height for the three cases. Here, the weak turbulence results have been shifted in time such that they rise above the PIC code noise level at $0.5 \tau_{ql}$ and have a noise baseline added. The time shifts are 2.8, 1.1 and 0.8 $\tau_{ql}$ respectively, and account for the higher effective thermal level due to noise in the PIC code.

In the cold ion case, the growth rates are very similar, as is the final level. A period of approximately exponential growth is seen between 0.5 and 1.5 $\tau_{ql}$, after which the instability saturates. {We therefore conclude that for cold ions, i.e. within the limits of WT theory, there is extraordinarily good agreement between the PIC and WT results,  not only for the peak Langmuir wavenumber, but also for the wave level and spectrum and, once noise has been accounted for, the wave growth rate. In other words, the quality of the PIC simulations is sufficient to reproduce the semi-analytical results, and moreover, the assumptions of the WT theory appear valid in this regime.}

The equal temperature case is shown for two independent runs of the PIC code, and illustrates the effects of the initial noise on the peak evolution. Small differences are seen, but by 2 $\tau_{ql}$ the levels are similar. However, the growth is notably slower than the weak-turbulence prediction, and in Figure \ref{fig:AllEnDens} we see that the daughter peak is shorter and broader than the cold ion case, which is not predicted by the WT result. Moreover the highest point of the peak is not where predicted, but is instead at larger wavenumber. {Thus it appears that once the ion-sound wave damping becomes significant, the WT theory no longer accurately describes the wave behaviour. This discrepancy is discussed further in the next section.}

\subsection{Interpretation as resonance broadening}

For the hot ions case, the discrepancy between PIC and weak-turbulence results in Figure \ref{fig:GrowthRates} becomes even more marked, as does the discrepancy in peak width and position in Figure \ref{fig:AllEnDens}. Due to computational limitations, and the longer box in this case, data is only available until 1.4$\tau_{ql}$, but by this stage the peak is well saturated. The level is far below the quasilinear prediction, and the width is much greater. 

\begin{figure}[]
\centering
\includegraphics[scale=0.5]{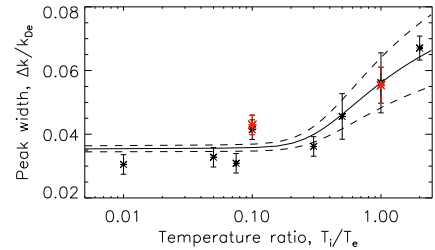}

\caption{$1/e$ widths for the daughter peak of Langmuir waves from the PIC code results against temperature ratio. The asterisks show the simulation results with error estimates given as the maximum of the time evolution around peak width saturation and the simulation wavenumber resolution. The solid and dashed lines are an estimate in the form $\Delta k_{ql}(1+A \gamma_s/\omega_s) $ with $\gamma_s$ given by Equation \ref{eq:isdamp}. The best fit by eye is for $A\simeq 3$ (solid line), while the dashed lines show $A\simeq 2$ (lower curve) and $A\simeq 4$ (upper curve). Second independent simulation results for $T_i=0.1T_e$ and $T_i=T_e$ are shown by red points.
 \label{fig:Rats}}
\end{figure}

This increase of peak width with ion-temperature suggests a link between the ion-sound wave damping rate and the peak width. To quantify the effect, we calculate the widths of the daughter wave peaks. The derived energy densities show noisy variations on scale similar to the simulation $k$ space resolution, and so were averaged over ten time bins and smoothed over $3 \textrm{d}k$ to reduce the effects of these variations. 
The $1/e$ peak widths were then calculated for each ion temperature for five time periods around the maximum height of the peak and the average taken.  The results are plotted in Figure \ref{fig:Rats}. Approximate error bars are calculated by taking the larger of $\pm \mathrm{d}k/2$ and the minimum and maximum values at the five times.


The effects of resonance broadening were discussed by Bian et al.\cite{2014JGRA..119.4239B} in the context of the beam plasma interaction, specifically its modification by scattering of electrons and/or plasma waves. In that case, for heavily damped Langmuir waves the delta function resonance in Equation \ref{eqn:4ql2} becomes broadened with approximate width $\Delta \omega$ the inverse lifetime of the wave. Here we extend this idea to the three-wave resonance between Langmuir and ion-sound waves, in which case the delta functions in Equations \ref{eqn:4ql_sSrc} and \ref{eqn:4ql_s} will be broadened with approximate width $\Delta \omega \simeq \gamma_s$. 

Because the ion-sound waves are low frequency this will only weakly affect the Langmuir wave frequency. However, the resultant ion-sound wavenumber is now constrained to only $\pm \Delta k_s /k_s  \simeq \Delta \omega_s/\omega_s$ (using $\omega_s \simeq k_s \mathrm{v}_s$). The participating Langmuir wavenumber, $k$ may therefore vary by this same amount, while still maintaining energy and momentum conservation for the interaction. We therefore expect a peak width of $\Delta k_{ql}(1+A \gamma_s/\omega_s) $ for $\Delta k_{ql}$ the peak width from the (unbroadened) weak turbulence simulations. The pre-factor $A$ is unknown, but expected to be of order unity as $\Delta \omega_s \simeq \gamma_s$ and therefore $\Delta k/k\simeq \Delta k_s /k_s  \simeq \gamma_s/\omega_s$

Figure \ref{fig:Rats} shows the widths derived from the simulations and this prediction for A between 2 and 4. The expected trend is clearly seen, with $A \simeq 3$. The simplicity of this model precludes an explanation for this value, but this is within the range expected. Moreover, the peak shapes shown in Figure \ref{fig:AllEnDens} are notably asymmetric and primarily extended towards larger wavenumber. This may be due to the limited wavenumber resolution affecting small wave numbers, or some unknown factor we do not predict. 
Figure \ref{fig:Rats} provides strong evidence that the analysis of the effect as resonance broadening is correct, and should certainly be studied in more detail in future. Moreover, it can be directly implemented in WT codes like the one used here, which may also yield interesting results.

\section{Discussion and conclusions}\label{sec:concl}

Quasilinear and weak turbulence theories have long been used to describe the evolution of a weak beam propagating in plasma. Such a scenario is of importance in several solar physics contexts, in particular for solar radio bursts. However, the weak turbulence theory is derived in the limit of cold ions and this is not always the case in such situations. We have therefore performed detailed numerical tests of WT theory using a series of very high particle number PIC simulations with EPOCH, for a weak Maxwellian beam in collisionless hydrogen plasma for a range of electron-ion temperature ratios. 

Convergence testing for the PIC code results was carried out for cold ions with particle numbers from 400 to 24000 particles per cell per species, including beam and background electrons and ions. The well known quasilinear plateau formation was observed for all numbers, but was seen to proceed much faster in the lower particle number cases, and for the smallest number to contain an erroneous high energy tail. We conclude that a few hundred ppc can reproduce plateau formation, but around 1000 ppc per species is needed to accurately reproduce the saturated state. Moreover, at least 10,000 ppc per species was needed to be able to correct for the super-thermal noise level and reproduce the time evolution. 

Fourier transforms of the electric field over a long time show clearly a well-defined Langmuir wave dispersion relation, identical to the usual Bohm-Gross prediction, and unaffected by a beam of the selected density ($n_b/n_e=10^{-4}$). A low frequency mode corresponding well to the plasma ion-sound mode was also seen, offering direct evidence for the expected resonant three-wave scattering processes.

Langmuir wave spectral energy densities were derived from the electric field, and compared to the WT results. For 4000 ppc and up, the primary Langmuir wave peak shape and position agreed well for all ion temperature cases. Moreover, detailed comparisons for 12000ppc in the cold ion case showed that the effects of ion-sound wave scattering were well reproduced, with a secondary peak appearing almost precisely where expected, and (once the initial noise was corrected for), showing the expected growth rate. 

For equal or double temperature ions the PIC results again match the quasilinear prediction for the initial Langmuir wave peak, and the beam relaxation. However, the secondary peak formed by three-wave scattering processes differs strongly from the theoretical expectation. In both cases it is broader and lower than expected, and has maximum at a larger wavenumber. 

Ten simulations over a range of ion temperatures were then performed, and show that the width of the scattered peak depends strongly on the electron-ion temperature ratio. Using the known ion-sound wave damping rate, we find that resonance broadening of the three-wave processes described in Section \ref{sec:QL_WT} can account for this. In this model the predicted quasilinear peak with is increased by a factor $(1+A \gamma_s/\omega_s) $ where $\gamma_s/\omega_s$ is the ion-sound wave damping rate in inverse wave periods, which depends on the electron-ion temperature ratio as shown in Equation \ref{eq:isdamp}, and $A$ is an unknown constant. A good fit between this model and the data was seen by eye for a value $A\simeq 3$. 

To conclude, we have tested the weak turbulence theory against PIC simulations, and found that in the cold ion limit there is very good agreement in electron and wave behaviour. However, when attempting to apply WT theory in plasma with ions approximately equal in temperature to the electrons, or hotter, there is significant deviation. We find a simple model which may account for this deviation, which can be implemented in WT codes such as used in this paper, and may yield very interesting results. Moreover, this correction strongly influences the Langmuir wave spectra in plasma of similar electron and ion temperature.  {This may be a vital consideration in, for example, the theory of Type III radio bursts, or the naturally enhanced ion-acoustic lines observed in the ionosphere, where the temperature ratio is close to 1 and the details of the wave spectrum are a vital factor.}

\begin{acknowledgements}
This work is supported by the European Research Council under the SeismoSun Research Project No. 321141 (HR, VMN),  STFC grant ST/L000733/1 (CSB), STFC Consolidated Grant ST/L000733/1 and the BK21 plus program through the National Research Foundation funded by the Ministry of Education of Korea (VMN).
We also thank the Warwick Undergraduate Research Support Scheme for support for a project contributing to this paper (MBCR). The EPOCH code was developed as part of the UK EPSRC funded projects EP/G054950/1 and EP/G056165/1. Computing facilities were provided by the Centre for Scientific Computing of the University of Warwick with support from the Science Research Investment Fund.
\end{acknowledgements}

\bibliography{QL_EPOCH}
\end{document}